\documentclass[aps,preprint,groupedaddress,superscriptaddress]{revtex4}

\usepackage{graphicx}
\usepackage{dcolumn}
\usepackage{bm}
\usepackage{color}

\begin{document}

\title{Metastability and topology in the magnetic topological insulator MnBi$_{2}$Te$_{4}$}%
\author{Jeonghwan Ahn,$^{1, \ast}$, Seoung-Hun Kang$^{1, \ast}$, Mina Yoon$^{1}$, Panchapakesan Ganesh,$^{2, \ddagger}$ Jaron T. Krogel$^{1, \dag}$}

\address{Materials Science and Technology Division, Oak Ridge National Laboratory, Oak Ridge, Tennessee 37831, USA}
\address{Center for Nanophase Materials Sciences, Oak Ridge National Laboratory, Oak Ridge, Tennessee 37831, USA}

\date{\today}

\begin{abstract}

 {We study the effect of stacking faults on the topological properties of the magnetic topological insulator MnBi$_{2}$Te$_{4}$ (MBT) using density functional theory calculations and the Hubbard $U$ being tuned with many-body diffusion Monte Carlo techniques. We show that a modest deviation from the equilibrium interlayer distance leads to a topological phase transition from a non-trivial to a trivial topology, suggesting that tuning the interlayer coupling by adjusting the interlayer distance alone can lead to different topological phases. Interestingly, due to the locally increased interlayer distance of the top layer, a metastable stacking fault in MBT leads to a nearly gapless state at the topmost layer due to charge redistribution as the topmost layer recedes. We further find evidence of spin-momentum locking in the surface state along with a weak preservation of the band inversion in the near gapless state, which is indicative of the non-trivial topological surface states for the metastable stacking fault. 
 Our findings provide a possible explanation for reconciling the long-standing puzzle of gapped and gapless states on MBT surfaces.
 }

\end{abstract}

\newpage

\maketitle

$^\ast$ These authors contributed equally to this work.

$^\dag$ krogeljt@ornl.gov
$^\ddag$ ganeshp@ornl.gov


\section*{\large Introduction}
\label{sec:introduction}

 MnBi$_{2}$Te$_{4}$ (MBT) has drawn a great deal of interest as the first realization of an intrinsic and stoichiometric topological insulator with $A$-type antiferromagnetic (AFM) order~\cite{otrokov2019prediction}. The structure of MBT is characterized by a layered structure consisting of septuple layers coupled together with van der Waals (vdW) interactions as illustrated in Fig.~\ref{fig:bulk}(a). Because of its crystal structure, there is a topological invariant under the product operation $S$ of time reversal ($\Theta$) and primitive lattice translation ($T_{1/2}$), namely $S = \Theta T_{1/2}$,
 giving rise to $Z_{2}$ topological classification~\cite{mong2010antiferromagnetic}. Consequently, gapped surface states are predicted to exist at the $S$-breaking surface of (0001), which makes MBT a highly desired platform as a single material to study an interplay between magnetism and topology. Broken time-reversal symmetry in MBT leads to novel topological phases such as quantum anomalous Hall effect~\cite{deng2020quantum,zhao2021even} and axion insulator~\cite{liu2020robust,lupke2022local}.
 
However, the existence of the gapped surface state has remained elusive in the sense that there has been no clear consensus among a series of experimental studies; the measured surface gaps differ depending on the sample~\cite{shikin2021sample,garnica2022native} and even gapless surface states have been reported~\cite{Hao_PhysRevX_2019,chen_2019_topological,li2019dirac,swatek2020gapless,nevola2020coexistence} since the first reports that support the gapped surface states of MBT~\cite{otrokov2019prediction}.
 Furthermore, a sizable spread among the measured surface gaps brought a fundamental question regarding whether the surface gap originated from the AFM order or not~\cite{chen_2019_topological}. 
 There have been many attempts to elucidate observed gapless and sample-dependent topological surface states, among which the different surface magnetic orderings from bulk ones like G-type or in-plane A-type AFM are predicted to result in linear dispersion~\cite{Hao_PhysRevX_2019,chen_2019_topological}. 
 Defects have been considered one of the probable sources to induce the surface magnetism distinct from the bulk one because of frequent observation of various defects in many experimental studies~\cite{zeugner2019chemical,huang2020native,lai2021defect,garnica2022native}. Especially, the Mn-involved defects are known to result in ferrimagnetic structure in the septuple layers~\cite{lai2021defect}, which are understood to shift the real-space localization of the topological surface states compared to the ideal case and lead to a substantial reduction of the surface gap~\cite{garnica2022native}.
 Besides, the topological properties were sensitive to the hydrostatic pressure (strain)~\cite{chen2019suppression,guo2021pressure,xu2022hydrostatic} because 
 structural and magnetic properties are found to be highly  coupled with pressure, reflecting their combined role in determining the topology of the system.
 Nevertheless, none of the above 
 fully resolves the ongoing debate on surface gaps.

For Te-based layered materials,
strong interlayer coupling due to interlayer hybridization is known to host correlated physics such as the metal-insulator transition between bilayer and monolayer IrTe$_{2}$ with the formation of charge density wave phase for monolayer~\cite{hwang2022large}, and Lifshitz transition for PtTe$_{2}$~\cite{lin2020dimensionality}. One can also expect substantial interlayer coupling in the MBT system because its valence band maximum (VBM) also mostly consists of p$_{z}$ orbitals. Hence interlayer hybridization can be maximized between p$_{z}$ orbitals aligned with the interlayer axis. More interestingly, since not only Te-p$_{z}$, but also Bi-p$_{z}$ orbitals are involved in the VBM of MBT, the spin-orbit coupling (SOC) effects would play some role in the interlayer hybridization between adjacent Te planes, which has not been the focus of the previous literature. The SOC-involved interlayer coupling would directly affect the band inversion and, thus, the topology of the system. Furthermore, the fact that the topology of MBT becomes nontrivial as the layers are stacked, but the single-layer MBT is topologically trivial~\cite{otrokov2019unique} underscores the interlayer coupling as a key factor to drive the non-trivial topology of the system.

We note that the interlayer coupling can be tuned by the nature of the local stacking, such as different stacking modes and twist angles, because of different resulting interlayer separations and the nearest neighbors across the vdW gap, which results in different physical properties. Indeed, according to a recent experiment ~\cite{yan2022perspective}, stacking faults induced by repeated bending of MBT bulk samples are observed to suppress the N{\'e}el temperature compared to the ideal case. The suppression effect receded after about one hour, is indicative of the metastability of the stacking faults induced by the bending.
Furthermore, it is remarkable to note that the impact of the stacking fault (or metastable stacking configuration) on the electronic structure, especially the topology, in MBT has not been explored yet though its existence is strongly suggestive in the mother material Bi$_{2}$Te$_{3}$~\cite{medlin2019unraveling} or MBT variant such as MnBi$_{4}$Te$_{7}$ or MnBi$_{6}$Te$_{10}$~\cite{yan2020type}.

The above facts motivated us to explore in greater depth the possible existence of a metastable stacking order in MBT and its influence on surface gap formation and corresponding topology. We firstly show that SOC and vdW interactions are intertwined due to interlayer hybridization involved with Bi p$_{z}$ and Te -p$_{z}$, supported by different predictions of interlayer binding energies and separations depending on the inclusion of SOC. Next, we demonstrate that modest expansion of the interlayer separations results in a topological phase transition, indicating the significance of the interlayer coupling to affect the topological phase, even with modest perturbations to the lattice. Furthermore, we show that a metastable local stacking mode in MBT characterized by a local increase in interlayer separation that lies right at the verge of creating inverted bands. As demonstrated in a 6-layer MBT system, we find that a metastable stacking at the MBT surface (which can be formed by sliding the topmost layer) leads to nearly gapless states localized at the topmost layer, consistent with the experimental observation of the gapless surface state for bulk MBT~\cite{lai2021defect,garnica2022native}. The spin is found to be locked to the momentum at the surface state and the gap closure at the surface corresponds to band inversion between the surface CBM and VBM, which is suggestive of its non-trivial topological surface states.

\section*{\large Results}
\label{sec:results_and_discussion}


\subsection*{Interplay between SOC and vdW interactions on interlayer hybridization}

\begin{figure}[h]
\centering
\includegraphics[width=6.5in]{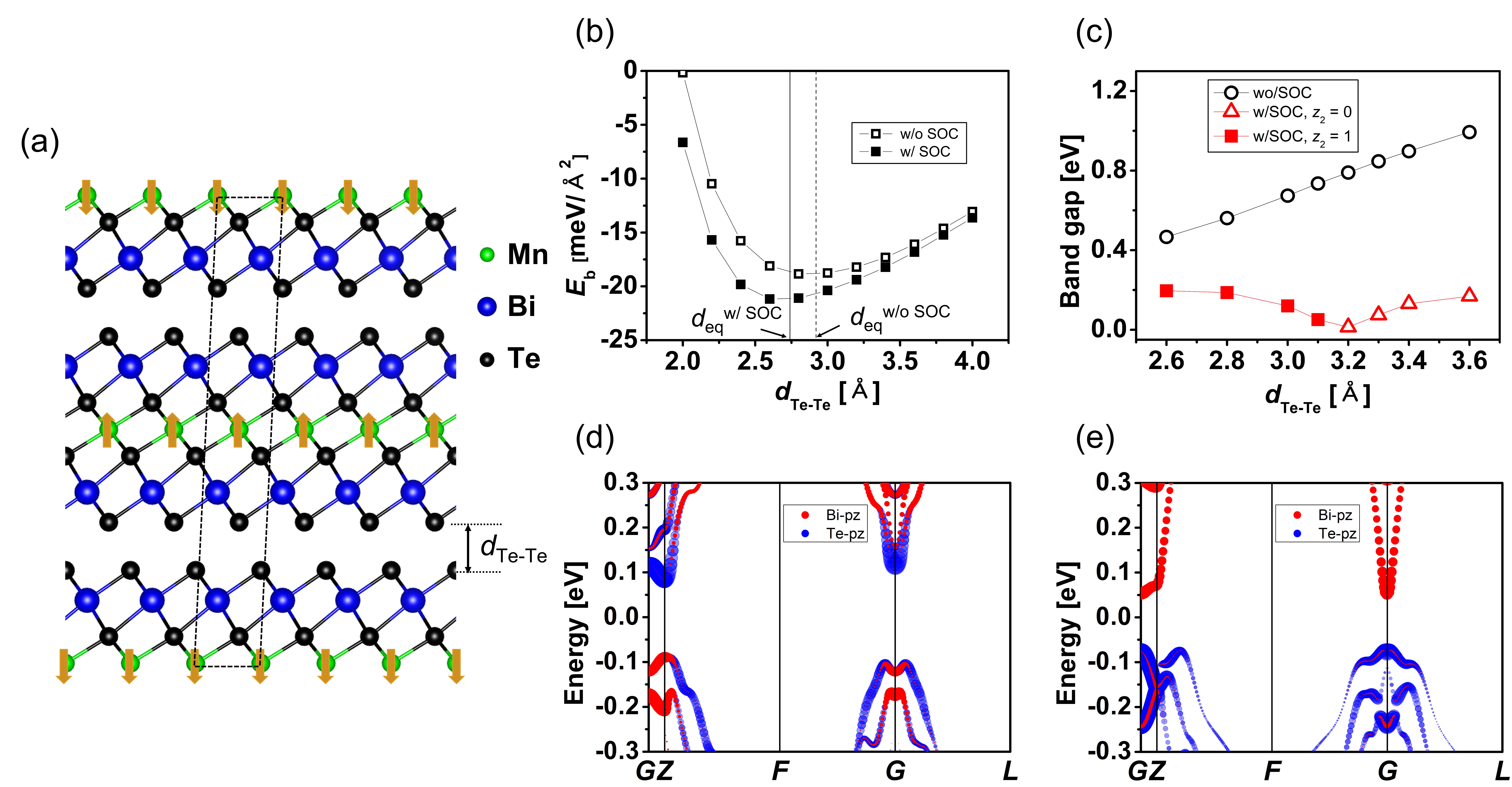}
\caption{(a) Crystal structure of MBT. The dotted lines represent its unit cell. (b) Interlayer binding energies for the bulk MBT as a function of the vertical distance between adjacent Te layers with/without SOC. The black solid (open) symbol represents interlayer binding energy with (without) SOC. (c) The Bandgap at $\Gamma$ of bulk MBT depends on including SOC as a function of the vertical distance between adjacent Te layers. The black open symbol represents the data without SOC, while the red solid (open) symbol denotes SOC data for which the corresponding topological $Z_{2}$ index is 1 (0). Band structures of the bulk MBT with the inclusion of SOC at (d) the equilibrium interlayer separation of 2.74~\AA~ and (e) 3.4 ~\AA. 
}
\label{fig:bulk}
\end{figure}

We first discuss the role of the SOC interaction in determining the structural and interlayer binding properties of the MBT layers. To this end, we present Fig.~\ref{fig:bulk}(b) to display interlayer binding energies of bulk MBT as a function of the vertical distance between adjacent Te layers, interlayer separation ($d_{\text{Te-Te}}$), with and without the inclusion of SOC at PBE+D3 level. The binding curves are found to be distinct depending on the inclusion of SOC. The equilibrium binding distance and interlayer separation for the SOC calculations are estimated to be 2.74~\AA~ and 21.3 meV/\AA$^{2}$, respectively. The predicted interlayer separation from the SOC calculations agrees with experimental values based on an X-ray measurement of 2.72(1)~\AA~\cite{zeugner2019chemical}. On the other hand, for the non-SOC calculations, the corresponding quantities are 2.92~\AA~ and 18.94 meV/\AA$^{2}$, which are smaller (longer) than the non-SOC binding energy (interlayer separation) about 11 $\%$ (7 $\%$). This discrepancy depending on the inclusion of SOC, indicates that SOC and vdW interactions are intertwined, which has not been the focus in previous literature. The intertwining of SOC and vdW interactions is found to be regardless of the inclusion of Hubbard U, the value of U, and different pseudopotentials.

In comparison with the interlayer binding curves of materials in a similar class as MBT, as shown in Supplementary Fig.1, the interlayer binding curves for the Bi-involved compounds display the difference between SOC and non-SOC calculations while the binding curves are identical for the Sb-involved compounds. This suggests that the Bi element possessing larger SOC strength than Sb is responsible for the intertwining SOC and vdW interactions. Mn does not play a significant role in the intertwining effect since binding curves for the Mn-involved compound show analogous features to those for non-magnetic counterparts. The difference in chalcogen atoms is sufficient to explain the different curvature of the interlayer binding curves. 


To further understand the interplay between SOC and vdW interactions, we investigated band structures and the band gap evolution as a function of $d_{\text{Te-Te}}$. Figure~\ref{fig:bulk}(d) displays band structures of the bulk MBT at the equilibrium interlayer separation when including SOC. Both orbital contributions of mainly Bi-p$_{z}$ and Te-p$_{z}$ to VBM signify the band inversion. Considering that this interlayer hybridization is aligned with the axis along which MBT septuple layers are stacked through the vdW interactions, the increasing SOC-involved interlayer hybridization as the septuple layers become close to each other is understood to result in the enhanced interlayer bindings in MBT when incorporating SOC. 
A trend of band gap as a function of interlayer separation shown in Fig.~\ref{fig:bulk}(c) is considered another clue indicating the coupling between SOC and vdW interactions in the interlayer hybridization. Figure~\ref{fig:bulk}(c) displays band gaps at $\Gamma$ point for different interlayer separations depending on the inclusion of SOC. When SOC is not included, the band gap is observed to be monotonically reduced with decreasing interlayer distances, while the band gap decreases as interlayer separation decreases but increases at the region of $d_{\text{Te-Te}} < 3.2$ in the presence of SOC. We attribute the SOC band gap trend distinct from the non-SOC case to the topological phase transition for the interlayer separations, which is verified with the $Z_{2}$ topological invariant changed from 1 to 0 between 3.1 $\sim$ 3.2~\AA~ as shown in Fig.~\ref{fig:bulk}(c). This is in line with the bulk gap closure between 3.1 $\sim$ 3.2~\AA~ and the band inversion disappears with increasing interlayer distance as shown in Fig.~\ref{fig:bulk}(e).
Interestingly, we notice that the band gap trend with SOC is identical to the band gap trend as a function of the SOC strength (See Fig.2D of Ref.~\cite{li2019intrinsic}) where phase transition occurs from the topologically non-trivial state to normal insulator. This suggests that interlayer separation controls the SOC strength through interlayer hybridization, and the modest deviation from its equilibrium value could result in a significant change in topological properties. From the above facts, we conclude that there is huge sensitivity in topological properties to the interlayer separation, and it is one of the most crucial factors responsible for the topological properties of MBT.

\noindent
\subsection*{Metastable stacking at the MBT bulk surface}

\begin{figure}
\centering
\includegraphics[width=6.2in]{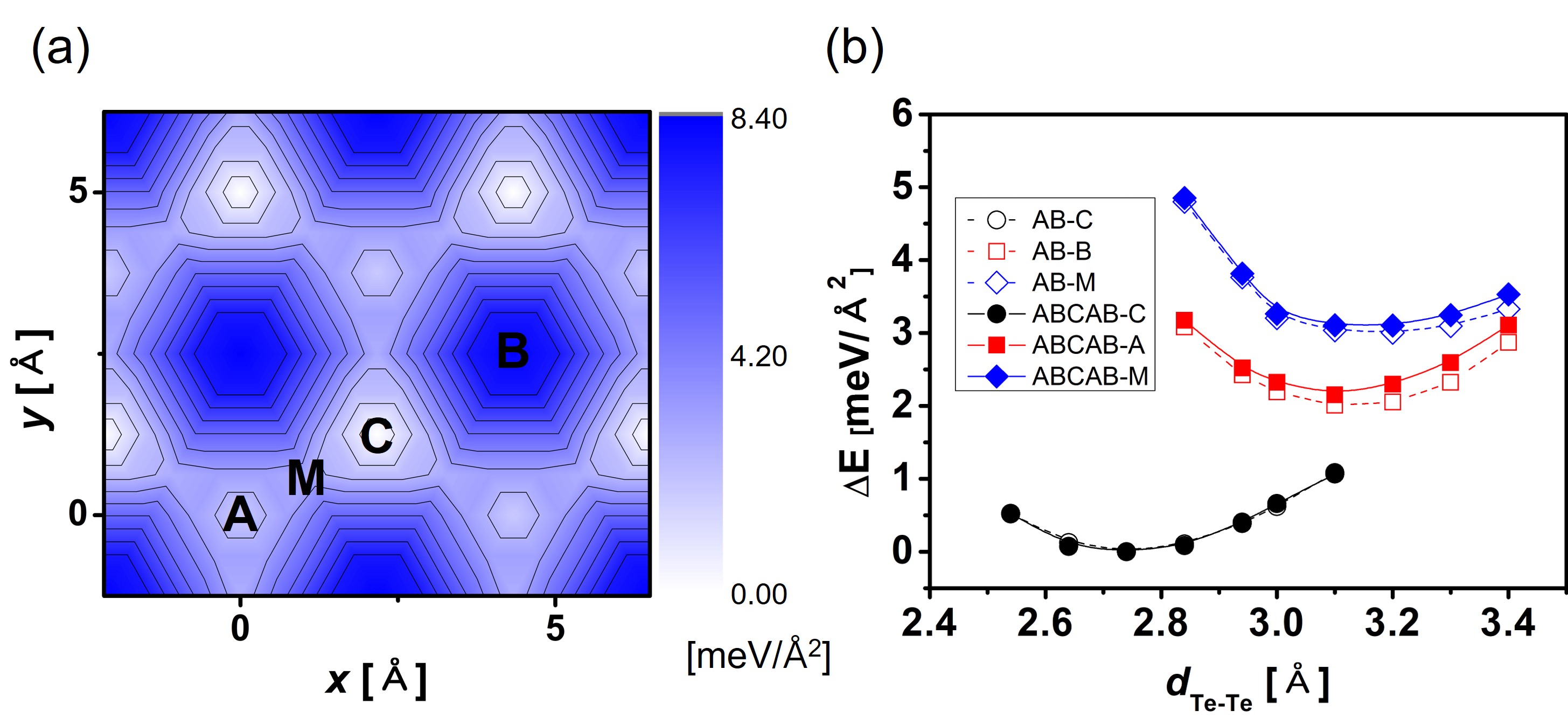}
\caption{(a) Potential energy surfaces for trilayer MBT with respect to sliding its topmost septuple layer with the inclusion of SOC. The letters A, B, C, and M denote the stacking position of the topmost septuple layer on top of the AB-stacked bilayer. (b) The DFT total energies as a function of interlayer separation between the adjacent Te plane of the topmost and second-topmost septuple layers, $d_{\text{Te-Te}}$, relative to the energy minima for the ABC- and ABCABC-stacked layers, respectively. Third-order polynomial fits are presented for a guide to the eye.}
\label{fig:PES}
\end{figure}

Considering that the topological properties of MBT might be tuned by stacking modes that have different interlayer separations from the ideal case based on the above results, we investigated the possible existence of metastable stacking mode for bulk MBT. To this end, we first computed the potential energy surface (PES) for trilayer MBT with respect to the sliding of its topmost layer on top of the underlying AB-stacked bilayer as displayed in Fig.~\ref{fig:PES}(a). 
For the consistency of the above binding curves for bulk MBT, the PES was computed with PBE+$U$(3.5 eV)+D3+SOC to reflect the intertwining of vdW and SOC interactions utilizing total energy curves as a function of interlayer separation between the topmost and second topmost layers at the given sliding with the thickness of MBT being fixed with the bulk value as shown in Fig.~\ref{fig:PES}(a). 
The letters A, B, and C in the contour map represent the relative stacking of the topmost layer above the AB-stacked one, and hence C appears to be the lowest-energy point because the ground-state stacking for the bulk structure is ABC. As shown in Fig.~\ref{fig:PES}(b), we could also identify the ABA stacking mode as a local minimum in the energy map with the energy difference from the ground-state structure being $\sim$ 2 meV/~\AA$^{2}$, which corresponds to about 11 $\%$ of the interlayer binding energy of the bulk MBT.  
We find from third-order polynomial fits that the equilibrium interlayer separation at ABA stacking has increased by about 11 $\%$ with an energy minimum residing at $d_{\text{Te-Te}}$ = 3.12~\AA{}, relative to ABC stacking whose minimum is located at $d_{\text{Te-Te}}$ = 2.74~\AA{} while the transition state (ABM stacking) exhibits the energy minimum at $d_{\text{Te-Te}}$ = 3.18~\AA{}.
Furthermore, one can clearly see in the PES energetically favored transition path from the global minimum to the local one (See path C-M-A Fig.~\ref{fig:PES}(a)) along which the corresponding barrier is estimated to be $\sim$ 3 meV/~\AA$^{2}$. This barrier and the small energy difference between global and local minima indicate that the metastable stacking mode might easily occur. 

Noting that 6-layer MBT is regarded as a sufficiently large slab model to approximate electronic structures of the surface of bulk MBT in the previous literature~\cite{shikin2021sample}, we examined with the 6-layer MBT whether metastable stacking of the topmost layer could occur at the surface of bulk MBT. Provided that the transition path for the 6-layer MBT, whose topmost layer is on top of the ABCAB-stacked one, is identical to that of the trilayer MBT, we compared their total energies relative to the lowest-energy stacking ones as a function of interlayer separation between the topmost and second topmost layers, $d_{\text{Te-Te}}$ when the topmost one is located at A, B or C as shown in Fig.~\ref{fig:PES}(b). The total energy curves of trilayer MBT are confirmed to be nearly identical to that of 6-layer MBT for all slidings, and consequently, energy differences between the stacking modes and corresponding interlayer separations become nearly identical to each other, with the equilibrium interlayer separation of the topmost layer residing at $d_{\text{Te-Te}}$ = 3.12~\AA. We also point out that slab and bulk calculations yield the same interlayer separations of 2.74~\AA~ for the lowest energy ABC and ABCABC stacking mode, suggesting that the interlayer separation is not sensitive to the number of MBT septuple layers because of its substantial thickness ($\sim 10.74$~\AA). This justifies our approach with a 6-layer MBT for energetic investigation of the possible metastable stacking mode at the bulk surface.
From this, we predict that these equilibrium quantities, along with the transition path to the metastable stacking mode, would be preserved with increasing thickness of MBT so that metastable stacking mode could also easily occur at the surface of bulk MBT with  a transition barrier of $\sim$ 3 meV/~\AA$^{2}$.

\begin{figure*}
\centering
\includegraphics[width=6.5in]{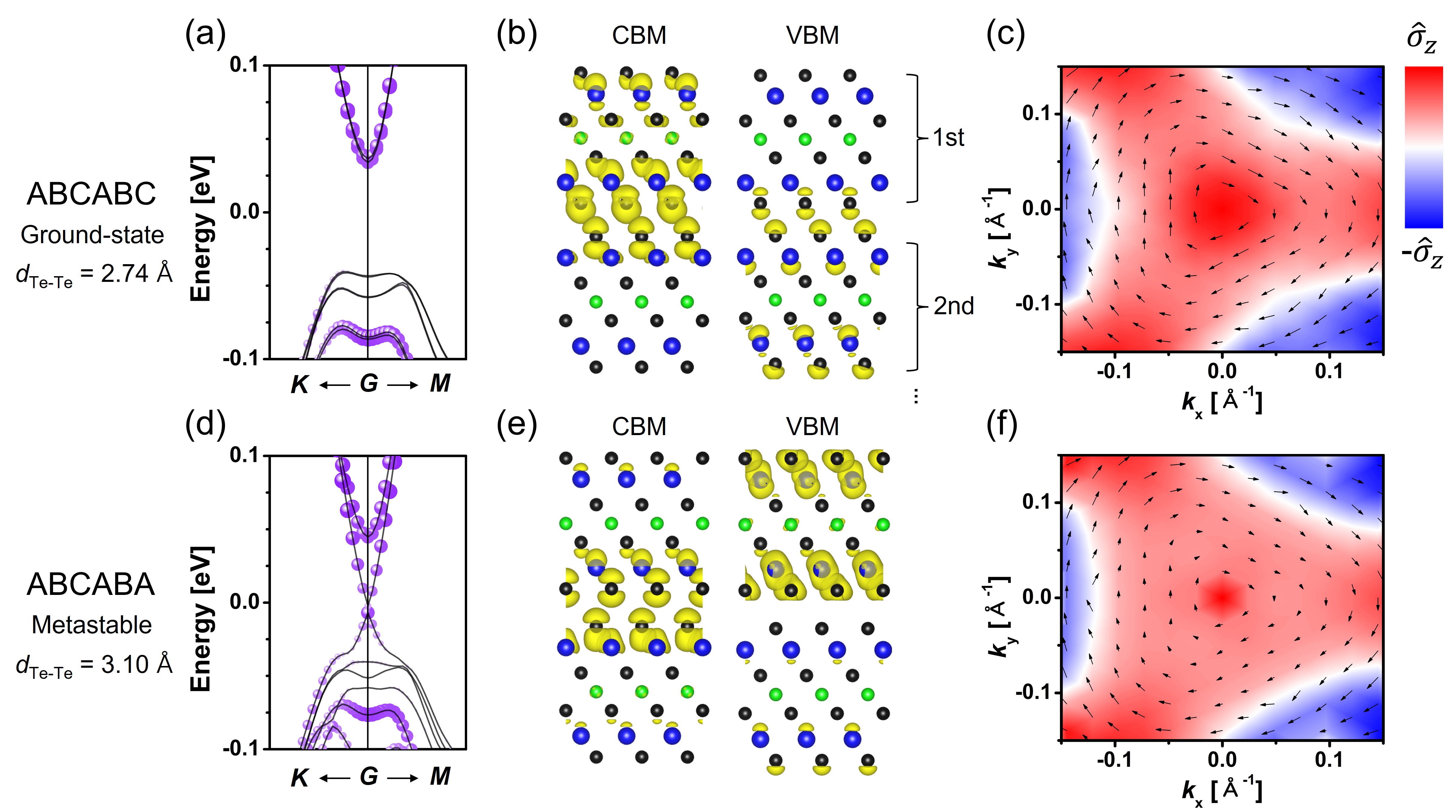}
\caption{Band structures for (a) the ground-state (ABCABC) and (d) metastable stacking (ABCABA) modes at their equilibrium. The solid violet symbols represent the weight of the projections of the Bloch states on the atoms belonging to the topmost and bottommost septuple layers. Spatial distributions of the CBM and VBM states at the gamma point are represented in (b) and (e) for the ground-state and the metastable stacking modes, respectively. The isosurfaces levels are $\pm 0.001$ $e$/bohr$^3$ and only two topmost layers are presented for clarity. 
Spin maps plotted in the reciprocal space for the CBM state (c) for the ground-state stacking and (f) metastable stacking modes. The size of black arrows and different colors represent the in-plane and out-of-plane spin components.} 
\label{fig:band_spin_texture}
\end{figure*}

\noindent
\subsection*{Electronic structure of metastable stacking mode}

We now present the band structures for the ground-state and metastable stacking modes identified in Fig.~\ref{fig:PES}. Fig.~\ref{fig:band_spin_texture}(a) presents the orbital projections of the surface atoms for the ground-state stacking at the equilibrium separation, defined here as those residing in the top- and bottom-most septuple layers, out of the total band structures denoted by violet symbols. It is found that occupied levels with strong surface contributions generally lie below sub-surface dominant states that are largely bulk-like, indicating that CBM and VBM originate from surface and bulk-like parts, respectively, as displayed in Fig.~\ref{fig:band_spin_texture}(b) showing the spatial distributions of CBM and VBM states at the gamma point. 
The surface gap of 118 meV is close to the experimentally measured gaps ranging from 20 $\sim$ 100 meV~\cite{otrokov2019prediction,lee2019spin,vidal2019surface,zeugner2019chemical,shikin2021sample,garnica2022native} while the bulk gap of 205 meV is found to be quantitatively in agreement with the experimental values ranging from 180 $\sim$ 220 meV~\cite{chen2019topological}. 
Furthermore, we present in Fig.~\ref{fig:band_spin_texture}(c) the spin projection onto the lateral plane of the reciprocal space for the surface-like CBM state, showing the spin is largely locked to the momentum, as is commonly observed in topologically non-trivial surface states. 
This qualitative observation is also consistent with the non-trivial topology of the bulk MBT at the equilibrium distance confirmed in Fig.~\ref{fig:bulk}(c).

On the other hand, the band structure of the metastable stacking mode is seen to be noticeably different compared with the ground-state one, as shown in Fig.~\ref{fig:band_spin_texture}(d); the surface-atom contributions appear in VBM as a result of the expanded separation distance of the topmost layer while CBM still remains as a surface band (See also Fig.~\ref{fig:band_spin_texture}(e)).
Consequently, the surface gap is estimated to be 5 meV at the metastable stacking mode, which may explain the occasional experimental findings of the gapless surface state~\cite{Hao_PhysRevX_2019,chen_2019_topological}.
We note that CBM for metastable stacking can still be seen as the surface-like state and the corresponding orbital character of is qualitatively analogous to that of the ground-state stacking whose surface state is in topologically non-trivial (See Fig.~\ref{fig:band_spin_texture}(b) and (e)) in the sense that the CBM state is mainly of Te-p$_{z}$ and Bi-p$_{z}$.
Furthermore, we also find spin-momentum locking for the metastable stacking mode, as shown in Fig.~\ref{fig:band_spin_texture}(f) through the corresponding signal is observed to be weakened compared to the case of ground-state stacking.
This suggests that the near gapless metastable stacking might still host a topologically non-trivial surface state even in the presence of weaker interlayer coupling.

\begin{figure}
\centering
\includegraphics[width=6.5in]{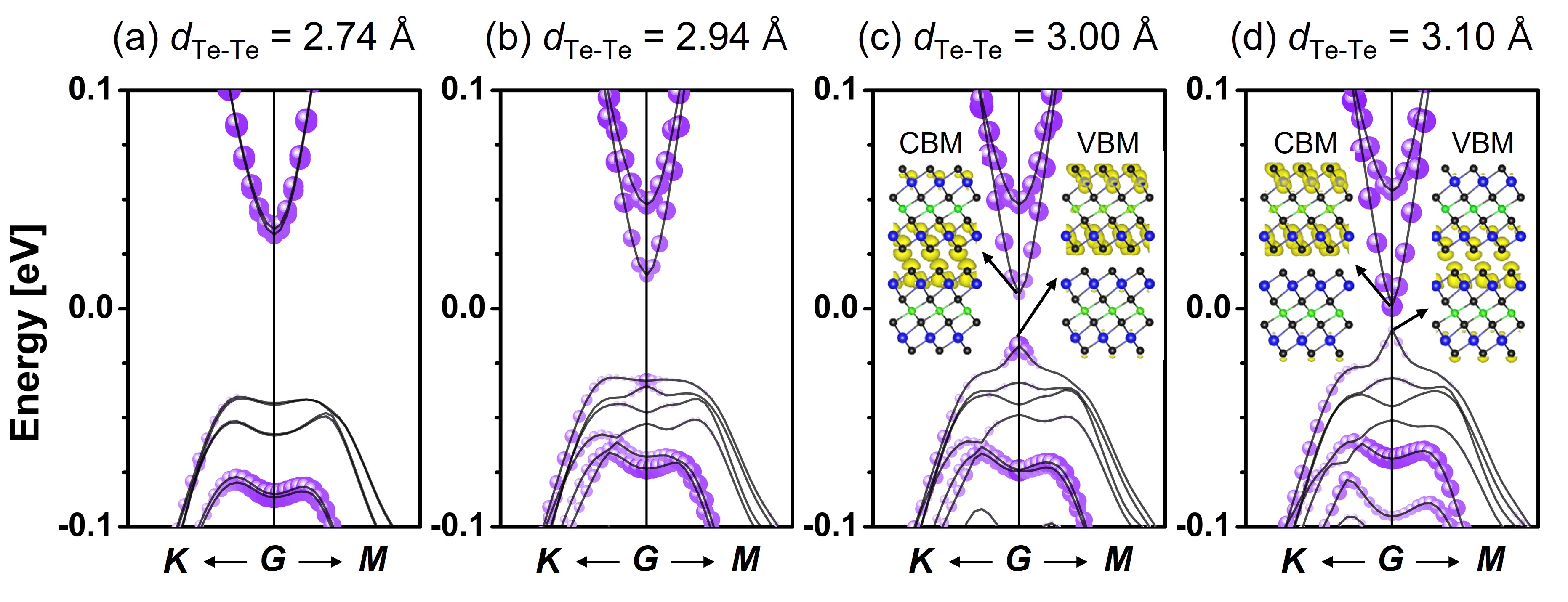}
\caption{Band structures of the ABCABC-stacked MBT layer for different $d_{\text{Te-Te}}$, that is, (a) $d_{\text{Te-Te}}$ = 2.74~\AA, (b) $d_{\text{Te-Te}}$ = 2.94~\AA, (c) $d_{\text{Te-Te}}$ = 3.0~\AA~ and (d) $d_{\text{Te-Te}}$ = 3.1~\AA. The solid violet symbols represent the weight of the projections of the Bloch states on the atoms belonging to the topmost and bottommost septuple layers. The insets represent Spatial distributions of the CBM and VBM states at the gamma point at each separation. The isosurfaces levels are $\pm 0.001$ $e$/bohr$^3$ and only two topmost layers are presented for clarity.}
\label{fig:6layer_band}
\end{figure}

Since the significant change in the band structure at the metastable stacking mode is understood to be due to the change of interlayer coupling by the expansion of the equilibrium separation at the topmost layer, we explore the effect of interlayer separation alone for the topmost layer on the band structure of the 6-layer MBT slab.
To this end, we analyze in Fig.~\ref{fig:6layer_band} an evolution of the band structures of the ground-state stacking mode by varying only interlayer separation of the topmost layer, $d_{\text{Te-Te}}$. 
As the topmost layer recedes, energy levels near the Fermi level significantly change. Especially, both VBM and CBM move toward the center of the gap, with the VBM character changing from bulk-like to surface-dominant while the CBM maintains strong surface contributions, giving rise to a significant band gap reduction of the surface state. 
Consequently, when the top-most layer reaches a separation of $d_{\text{Te-Te}}$ = 3.1~\AA~(See Fig.~\ref{fig:6layer_band}(e)), both CBM and VBM mainly have the surface contributions and the surface gap is estimated to be 11 meV as similar to the case of the metastable stacking configuration in Fig.~\ref{fig:band_spin_texture}(d). 
This indicates that the surface states (gap) are very sensitive to the interlayer separation between the topmost and second-topmost layers, which could be considered a crucial factor in explaining gapless or sample-dependent surface gap measured in the previous experiments~\cite{shikin2021sample,garnica2022native,Hao_PhysRevX_2019,chen_2019_topological}.

As is clearly seen in the spatial distributions of CBM and VBM at the gamma point presented in the inset of Fig.~\ref{fig:6layer_band}, the CBM state remains localized at the interface of the topmost layer, while the localization of the VBM state changes from the bulk to the topmost region as the topmost layer recedes. Thus the gapless state, if any, is expected to exist primarily at the topmost layer exposed to the spectroscopic measurement's light source.
Furthermore, we notice the band inversion occurring between $d_{\text{Te-Te}}$ = 3.0 and 3.1~\AA~ since the spatial distributions of CBM and VBM are observed to be switched. 
Analysis of the orbital projections reveals that VBM and CBM mainly have Te-p$_{z}$ and Bi-p$_{z}$ character, respectively, at $d_{\text{Te-Te}}$ = 3.0 and 3.1~\AA~ and these are seen to be inverted at $d_{\text{Te-Te}}$ = 3.1~\AA. 
The separation range of the disappearance of band inversion, which coincides with the one for the band-gap closure, might suggest a topological phase transition occurring between $d_{\text{Te-Te}}$ = 3.0 and 3.1~\AA. This requires band-gap closure at the intermediate critical point, which is consistent with the gapless surface state~\cite{lai2021defect,garnica2022native}. 
We further see that the splitting in the VBM states is about 65 meV near $d_{\text{Te-Te}}$ = 3.1~\AA{} in the ABCABC stacking, reproducing well the observed 70 meV $k_z$ dispersion in the VBM states for the MBT system with a gapless surface state \cite{Hao_PhysRevX_2019}.
In addition, it is worthwhile to note that the total band structures of ABCABA and ABCABC at $d_{\text{Te-Te}}$ = 3.1~\AA{} are found to be nearly identical to each other, but the band is inverted at the topmost layer for the ABCABA stacking while there is no band inversion for the ABCABC one (See Fig.~\ref{fig:band_spin_texture}(e) and Fig.~\ref{fig:6layer_band}(d)). This suggests that changes in interlayer hybridization in the vicinity of $d_{\text{Te-Te}}$ = 3.1~\AA~ for the two stackings are sufficient to flip the band inversion when it is close to forming; however, the gap closing is clearly dominated by changes in interlayer coupling driven by the separation distance, underscoring the significance of the interlayer coupling in understanding the surface gap closure and its corresponding topological phase.

\section*{\large Discussion}
\label{sec:discussion}

Combined, our results suggest that local realizations of this metastable stacking mode in MBT result in near-gap closings, which leads us to conclude that the proposed metastable stacking mode could explain the controversial spread in the experimentally measured surface gap. It is worthwhile to note that tuning interlayer separation was recently achieved through an electrochemical protocol in MoS$_{2}$ by introducing active sites of Ce ions and intercalating Pt atoms in the interlayer region~\cite{ding2021bidirectional}. This makes us expect the interlayer separation of MBT to be also tuned in a similar fashion, which opens a possibility to manipulate its topological phase via direct control of interlayer separation.


We have proposed the metastable stacking mode that could consistently explain the spread in the experimentally measured surface gap with the non-trivial topology.
Especially in light of our results, the surface-gap closure may be understood as a part of a topological phase transition induced by the expansion of the interlayer separation at a metastable stacking configuration. 
The interplay between the interlayer hybridization of the p$_{z}$ orbitals of Bi and Te and the effective spin-orbit coupling strength is important to this effect. This reveals the crucial role of interlayer separation as the tuning factor of interlayer coupling, which directly impacts the energetics, band structure, and the topology of MBT. In this sense, we point out that increased interlayer separation could be induced by surface defects~\cite{zeugner2019chemical,huang2020native,lai2021defect,garnica2022native} or intercalation~\cite{mazza2022surface}, which were proposed to elucidate the gapless surface state in the previous literature. Thus, our study provides deeper insight into understanding the interplay between structure and topology while shedding light on the possibility of manipulating 
the topological phase via engineering the interlayer separations.

\section*{\large Methods}
\label{sec:methods}

The DFT calculations were implemented with the QUANTUM ESPRESSO package~\cite{giannozzi09}. To solve the Kohn-Sham equations, we used optimized norm-conserving Vanderbilt (ONCV) pseudopotentials~\cite{hamann2013} for the atomic species in MBT. The plane-wave cutoff was set to be 120 Ry, and the $15 \times 15 \times 3$ ($15 \times 15 \times 1$) Monkhorst–Pack~\cite{monkhorst76} $k$ meshes were used for the bulk (few-layer) MBT. The structural relaxations were done with the accuracy of 10$^{-4}$ Ry/Bohr. The van der Waals interactions were included with DFT+D3 method~\cite{grimme10}, and the Hubbard U~\cite{dudarev98,liechtenstein1995density} was applied to properly describe the Mn-driven magnetic states. The spin-orbit interactions were also included to discuss the effects of spin-orbit coupling (SOC) on the electronic and structural properties and the energetics at the DFT level. In structural models of few-layer MBT systems, we used a vacuum distance of 60~\AA~along the vertical direction up to the trilayer while 120~\AA~is used for 6-layer MBT to minimize spurious interactions between periodic images. From the DFT results, we obtain the maximally localized Wannier functions for 4$s$- and 3$d$-orbitals of Mn atom, 6$p$ of Bi atom and 5$p$-orbitals on Te atoms by using the WANNIER90 code~\cite{MOSTOFI2014}. These were used to analyze the topological properties. The topological index are calculated by the WANNIERTOOLS~\cite{WU2018}, which is based on the Wannier charge center calculation method~\cite{AS2011}.

We utilized QMCPACK code for the fixed-node DMC calculations to tune Hubbard U for the monolayer MBT as in our previous study for the bulk MBT~\cite{bennett2022magnetic}. The pseudopotentials which were generated for Mn~\cite{annaberdiyev2018new} or tested for Bi~\cite{metz2000small} and Te~\cite{stoll2002relativistic} within the ccECP scheme~\cite{bennett2017new,annaberdiyev2018new}, yielded bulk Mn magnetization in benchmark quality. Slater-Jastrow type trial wavefunctions were used with the Jastrow parameters up to three-body ones to incorporate electron-ion, electron-electron, and electron-electron-ion correlations, which were optimized with variational Monte Carlo calculations based on the linear method of Umrigar {\it et al.}~\cite{umrigar07}. In subsequent DMC calculations, we used a time step of $\tau = 0.005$ Ha$^{-1}$ and size-consistent T-moves for variational evaluation of non-local pseudopotentials when imaginary time projection proceeds~\cite{kim18,kent20}.
 
 To optimize the nodal surface of the many-body wavefunctions for the monolayer MBT, we computed its DMC total energies for the 7-atom ferromagnetic cell as a function of U, for which the PBE+U orbitals are used in constructing the wavefunctions. Fitting DMC energies to the third-order polynomial fit results in the optimal value of U= 3.7(3) eV (See Supplementary Fig.2). The DMC-optimized U for the monolayer is observed to be identical to the corresponding value for the bulk MBT within the statistical error bar~\cite{bennett2022magnetic}, which provides the onsite Mn magnetization that is in good agreement with neutron diffraction measurement~\cite{Ding_PhysRevB_2020}. 
 The nearly identical U values for bulk and monolayer MBT are understood by the similar chemical environment around Mn atoms, even though layer thickness varies. 
 Based on this fact, the following DFT calculations for few-layer MBT were performed with the DMC-informed Hubbard U for the bulk value of 3.5 eV for the consistency of our calculations dealing with both few-layer and bulk MBT.


\section*{\large Acknowledgments}

Work performed by J.A., P.G., and J.T.K. (DFT calculations, original idea) was supported by the U.S. Department of Energy, Office of Science, Basic Energy Sciences, Materials Sciences and Engineering Division, as part of the Computational Materials Sciences Program and Center for Predictive Simulation of Functional Materials.
Work performed by S.-H. K. and M. Y. (topological index and spin texture calculations) was supported by the U.S. Department of Energy (DOE), Office of Science, National Quantum Information Science Research Centers, Quantum Science Center. 
An award of computer time was provided by the Innovative and Novel Computational Impact
on Theory and Experiment (INCITE) program. This research used resources of the Oak Ridge Leadership Computing Facility, which is a DOE Office of Science User Facility supported under Contract DE-AC05-00OR22725.  This research also utilized resources of the National Energy Research Scientific Computing Center, a DOE Office of Science User Facility supported by the Office of Science of the U.S. Department of Energy under Contract No. DE-AC02-05CH11231 using NERSC award BES-ERCAP0024568.

\section*{\large Data Availability}
The data that support the findings of this study are available
within this article, its supplementary material, and in the materials
data facility\cite{Blaiszik2016, Blaiszik2019} at [link to be provided upon acceptance of this manuscript].

\section*{\large Competing interests}
The authors declare no competing interests.

\section*{\large References}
\bibliography{arXiv.bbl}

\end{document}